%% file: acl_latex.tex
\pdfoutput=1

\documentclass[11pt]{article}

\usepackage[]{acl}

\usepackage{times}
\usepackage{latexsym}

\usepackage{times}
\usepackage{latexsym}

\usepackage{times}
\usepackage{latexsym}

\usepackage[normalem]{ulem}
\usepackage{mathrsfs}
\usepackage{amsthm}
\usepackage{amsmath}
\usepackage{amssymb, bm}
\usepackage{multirow}
\usepackage{booktabs}
\usepackage{enumitem}
\usepackage{times}
\usepackage{latexsym}
\usepackage{array}
\usepackage{multirow}
\usepackage{graphicx}
\usepackage{amsfonts}
\usepackage{subcaption}
\usepackage{xspace}
\usepackage{tabularx}
\usepackage{algorithm}
\usepackage{arydshln}
\usepackage{algpseudocode}
\usepackage{colortbl}
\usepackage{tcolorbox}

\definecolor{mypurple}{RGB}{111,61,121}
\definecolor{myblue}{RGB}{46,88,180}
\definecolor{myred}{RGB}{181,68,106}
\definecolor{textorange}{RGB}{237,125,49}
\definecolor{textblue}{RGB}{46,117,181}
\definecolor{textgreen}{RGB}{112,173,71}

\newcommand{\ours}{CSQE\xspace}
\newcommand{\red}[1]{{\color{myred}{{#1}}}}
\newcommand{\green}[1]{{\color{textgreen}{{#1}}}}

\usepackage[T1]{fontenc}

\usepackage[utf8]{inputenc}

\usepackage{microtype}

\usepackage{inconsolata}

\title{Corpus-Steered Query Expansion with Large Language Models}

\author{Yibin Lei\textsuperscript{1}, Yu Cao\textsuperscript{2}, Tianyi Zhou\textsuperscript{3}, Tao Shen\textsuperscript{4}, Andrew Yates\textsuperscript{1}\\
\textsuperscript{1}{University of Amsterdam} \quad \textsuperscript{2}{Tencent IEG} \\
\textsuperscript{3}{University of Maryland}
\quad \textsuperscript{4}{University of Technology Sydney} \\
\texttt{\{y.lei, a.c.yates\}@uva.nl}, \texttt{rainyucao@tencent.com} \\
\texttt{tao.shen@uts.edu.au},
\texttt{tianyi@umd.edu}
}

\begin{document}
\maketitle

\input{abstract.tex}
\input{intro.tex}

\input{method.tex}
\input{experiments.tex}

\input{analysis.tex}
\input{conclusion.tex}
\input{limitations.tex}
\input{acknowledgement.tex}
\bibliography{anthology,custom}

\input{appendix.tex}

\end{document}

%% file: abstract.tex
\begin{abstract}
Recent studies demonstrate that query expansions generated by large language models (LLMs) can considerably enhance information retrieval systems by generating hypothetical documents that answer the queries as expansions.
However, challenges arise from misalignments between the expansions and the retrieval corpus, resulting in issues like hallucinations and outdated information due to the limited intrinsic knowledge of LLMs.
Inspired by Pseudo Relevance Feedback (PRF), we introduce Corpus-Steered Query Expansion (CSQE) to promote the incorporation of knowledge embedded within the corpus.
CSQE utilizes the relevance assessing capability of LLMs to systematically identify pivotal sentences in the initially-retrieved documents. These corpus-originated texts are subsequently used to expand the query together with LLM-knowledge empowered expansions, improving the relevance prediction between the query and the target documents. Extensive experiments reveal that CSQE exhibits strong performance without necessitating any training, especially with queries for which LLMs lack knowledge.\footnote{Our code is publicly available at \url{https://github.com/Yibin-Lei/CSQE}.}

\end{abstract}

%% file: intro.tex
\section{Introduction}

Query expansion enhances the effectiveness of information retrieval systems by incorporating additional texts into the original query, which are traditionally identified via pseudo-relevance feedback~\cite{prf1, prf2} or by leveraging external lexical knowledge sources~\cite{lexical_expansion1, lexical_expansion_2}. Recent studies~\cite{{hyde, query2doc, prompt_qe, generative_relevance_feedback}} show query expansions generated by LLMs are able to significantly boost retrieval effectiveness, especially in zero-shot scenarios. For instance, \citet{hyde} demonstrates the effectiveness of utilizing LLMs to generate hypothetical documents answering the original query as additional texts to augment the query. \citet{generative_relevance_feedback}
show the efficacy of applying pseudo-relevance feedback upon the LLM-generated answers for expansion. Despite variations in prompts or expansion methods, a common foundational element across these approaches is the reliance on the intrinsic knowledge of LLMs.

\begin{figure}[t]
\setlength{\abovecaptionskip}{0.1cm}
\setlength{\belowcaptionskip}{-0.3cm}
    \centering
    \includegraphics[width=0.85\linewidth]{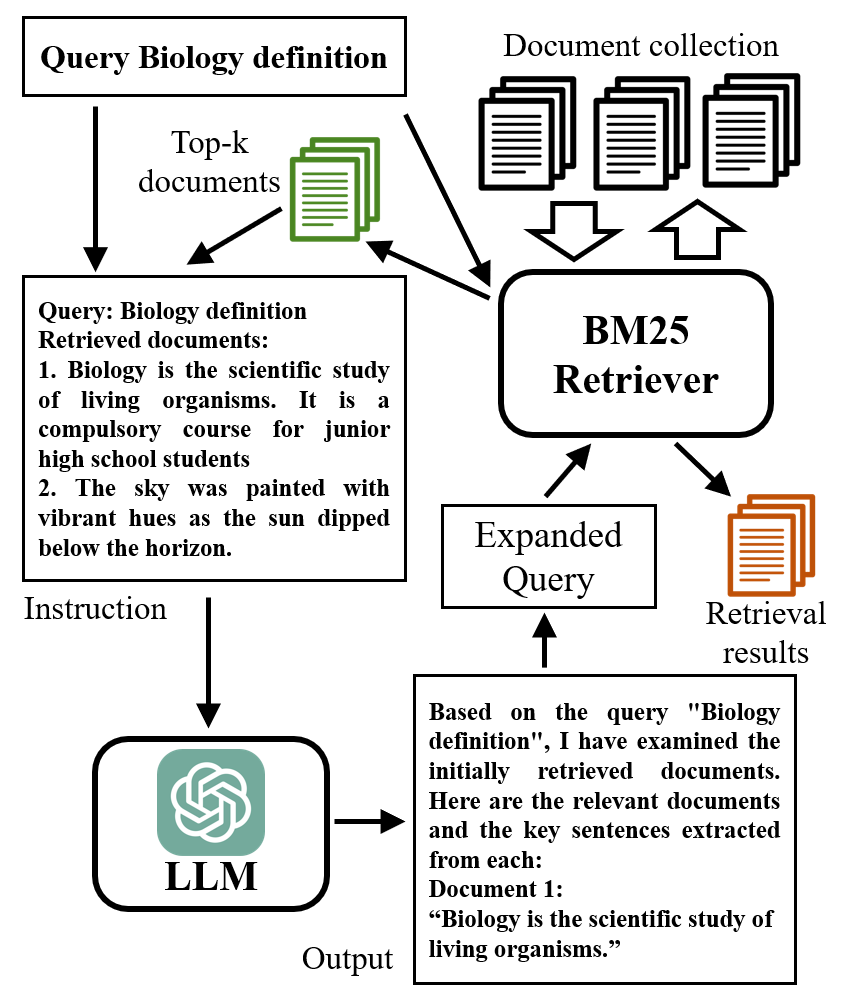}
    \caption{Overview of \ours. Given a query \textit{Biology definition} and the top-2 retrieved documents, CSQE utilizes an LLM to identify relevant document 1 and extract the key sentences from document 1 that contribute to the relevance. The query is then expanded by both these corpus-originated texts and LLM-knowledge empowered expansions (i.e., hypothetical documents that answer the query) to obtain the final results.}
    \label{fig:illustration}
\end{figure}

Despite their effectiveness, generations that rely on the intrinsic parametric knowledge within LLMs encounter various issues. These include hallucination~\cite{hallucination_1}, inability to update~\cite{timeqa}, and a deficiency in long-tail knowledge~\cite{long-tail}. Such generations may introduce irrelevant or misleading texts, degrading retrieval performance~\cite{expan_fail}. These query expansions can be seen as an evolution of earlier query expansions reliant on external lexical knowledge. In contrast, tradition PRF that typically chooses additional texts from the top-retrieved documents, has received less attention. However, given that the expanded texts are sourced directly from the original documents, these methods hold significant potential for enhancing factuality.

To this end, we propose Corpus-Steered
Query Expansion (CSQE). Unlike methods that rely on the intrinsic parametric knowledge of LLMs, CSQE exclusively leverages the strong relevance assessing capability of LLMs~\cite{relevance_assess_1, relevance_assess_2}.
As illustrated in Figure~\ref{fig:illustration}, given a query and its initially retrieved documents, CSQE utilizes a LLM to first identify relevant documents to the query and then extracts pivotal sentences that contribute to their relevance. These corpus-originated texts are then combined together with LLM-knowledge empowered expansions to expand the original query.
By incorporating query expansions that strictly originate from the corpus, CSQE balances out the limitations commonly found in LLM-knowledge empowered expansions.

To sum up, our contributions are 3-fold:

\noindent 1) We propose CSQE, which exclusively exploits the relevance assessing capability of LLMs to overcome the hinderance posed by LLM-knowledge empowered expansions.

\noindent 2) Experimental results reveal that \ours combined with a simple BM25 model, without necessitating any training, outperform not only LLM-knowledge empowered expansion methods but also the SOTA supervised Contriever$^\text{FT}$ model across two high-resource web search datasets and six low-resource BEIR datasets.

\noindent 3) Further analysis demonstrates the advantages of BM25 over dense retrieval with query expansion from LLMs, as well as query expansion over large-scale fine-tuning upon Contriever.

%% file: method.tex
\section{Method}
In this section, we first describe how we implement a Knowledge Empowered Query Expansion baseline based on LLMs (KEQE), then present the details of \ours to enhance BM25. 

\paragraph{KEQE}
Inspired by recent works that directly generate hypothetical documents to answer the query via LLMs for boosting retrieval~\cite{hyde, query2doc, prompt_qe, generative_relevance_feedback}, we implement a KEQE baseline in a similar pattern for fair comparison. Given a query $q$, we use LLMs to generate the hypothetical answer $a$ via a task-agnostic prompt shown in Table~\ref{tab:ge_prompt}. The concatenation of $q$ and $a$ is then used as the expanded query to BM25 to retrieve the final results.

It is worth noting that these hypothetical documents are inevitably susceptible to issues like hallucination that can adversely affect retrieval performance, due to the limitation of LLMs' intrinsic knowledge. To mitigate such problems, we propose CSQE to incorporate corpus-originated expansions with knowledge embedded in the corpus.

\begin{table}[htbp]
\setlength{\abovecaptionskip}{0.2cm}
\setlength{\belowcaptionskip}{-0.0cm}
  \tiny
  \centering
  \begin{tabularx}{\linewidth}{|X|}
    \toprule
    \textbf{KEQE Prompt} \\
    \midrule
Please write a passage to answer the question

Question: \{$q$\}

Passage:\\
    \bottomrule
  \end{tabularx}
  \caption{Prompt of KEQE. $\{\cdot\}$ denotes the placeholder for the corresponding text. }
  \label{tab:ge_prompt}
\end{table}

\paragraph{\ours}
Given a query $q$ and the document collection $\mathcal{D}$, we first retrieve the top-$k$ documents $\{d_1, d_2, \ldots, d_k\}$ using BM25. Then we elicit large language models to directly perform pseudo-relevance feedback via one-shot prompting as shown in Table~\ref{tab:prf_prompt}, where the current retrieved documents are integrated.
The learning context in the prompt is constructed from the TREC DL19 dataset for constraining the structure of generated texts. Noting that such a prompt remains unchanged for all tasks, we can therefore consider our method with minimal relevance supervision and being a zero-shot approach for all datasets excluding DL19 (which is used in the prompt).

Based on the above prompting, the generation of LLMs will contain (1) the indices of documents that are identified as relevant to the query and (2) the key sentences that contribute to their relevance, denoted as $S=\{s_1, s_2, \ldots, s_n\}$. Then we expand the query by concatenating $q$, all sentences in $S$, and the generations from KEQE to form a new query for BM25 retrieval, where the results in this turn are regarded as the final retrieved documents. 
Since these key sentences are usually identical to the existing texts in the corpus\footnote{In our preliminary study, we found 830 out of 1000 key sentences extracted by GPT-3.5-Turbo are identical to sentences in the initially-retrieved documents.},
they are much less prone to issues such as hallucinations and shortness of long-tail knowledge and can balance out the limitations of KEQE expansions.

To increase diversity, we sample $N$ generations from the LLM for expansion. For KEQE, $N=5$. As CSQE involves both KEQE and corpus-originated expansions, we sample $N=2$ for both KEQE and corpus-originated expansions, in total only 4 generations for fair comparison. We repeat the initial query $q$ a number of times equal to the number of expansions during concatenation.

\begin{table}[h]
\setlength{\abovecaptionskip}{0.1cm}
\setlength{\belowcaptionskip}{-0.3cm}
  \tiny
  \centering
  \begin{tabularx}{\linewidth}{|X|}
    \toprule
    \textbf{CSQE Prompt} \\
    \midrule

Query: "how are some sharks warm blooded"

Retrieved documents:

1. Most sharks are cold-blooded. Some, like the Mako and the Great white shark, are partially warmblooded (they are endotherms)\ldots

2. Are sharks cold-blooded or warm-blooded? Sharks have a reputation as cold-blooded and despite how negative that term is\ldots

3. Great white sharks are some of the only warm blooded sharks. This allows them to swim in colder waters in addition to warm, tropical waters\ldots

You will begin by examining the initially retrieved documents and identifying the ones that are relevant, even partially, to the query. Once the relevant documents are identified, you will extract the key sentences from each document that contribute to their relevance.
\\ \\
Based on the query "how are some sharks warm blooded", I have examined the initially retrieved documents. Here are the relevant documents and the key sentences extracted from each: 

Document 1:\\
"Most sharks are cold-blooded. Some, like the Mako and the Great white shark, are partially warm-blooded (they are endotherms)."

Document 3:\\
"Great white sharks are some of the only warm-blooded sharks."\\ \\

Query: "$\{q\}$"

Retrieved documents:

1. $\{d_1\}$

2. $\{d_2\}$

\ldots

$\{k\}$. $\{d_k\}$

You will begin by examining the initially retrieved documents and identifying the ones that are relevant, even partially, to the query. Once the relevant documents are identified, you will extract the key sentences from each document that contribute to their relevance.\\
    \bottomrule
  \end{tabularx}
  \caption{Prompt of \ours. $\{\cdot\}$ denotes the placeholder for the corresponding text. Refer to Appendix~\ref{instruc:prf-llm} for the complete prompt.}
  \label{tab:prf_prompt}
\end{table}

\begin{table*}[h]
\setlength{\abovecaptionskip}{0.2cm}
\setlength{\belowcaptionskip}{-0.0cm}
\centering
\small
\begin{tabular}{l|ccc|ccc}
\toprule
 & \multicolumn{3}{c|}{DL19}          & \multicolumn{3}{c}{DL20} \\
\midrule
 & mAP  & nDCG@10 & Recall@1k & mAP  & nDCG@10 & Recall@1k \\
\midrule
\multicolumn{7}{l}{\textit{w/o training}} \\
BM25 & 30.1 & 50.6 & 75.0 & 28.6 & 48.0 & 78.6 \\
BM25+RM3 & 34.2 & 52.2 & 81.4 & 30.1 & 49.0 & 82.4  \\
Contriever+HyDE & 41.8 & 61.3 & 88.0	& 38.2	& 57.9 & 84.4 \\
BM25+GRF  & 44.1 & 62.0 & 79.7 & \textbf{48.6} & 60.7 & 87.9 \\
BM25+Q2D/PRF & 43.6 &	65.4 & 87.1 & 40.5	& 61.0	& 87.2\\
BM25+KEQE & 45.0 & 65.9 & \textbf{88.8} & 42.8 & 60.5	& 88.3  \\
BM25+CSQE & \textbf{47.2} & \textbf{67.3} & 88.5	& 46.5	& \textbf{66.2} & \textbf{89.1} \\
\midrule
\multicolumn{7}{l}{\textit{reference. w/ training}}\\
DPR  & 36.5 & 62.2 & 76.9 & 41.8 & 65.3 & 81.4 \\
ANCE & 37.1 & 64.5 & 75.5 & 40.8 & 64.6 & 77.6 \\
Contriever$^\text{FT}$ & 41.7 & 62.1 & 83.6 & 43.6 & 63.2 & 85.8 \\
\bottomrule
\end{tabular}
\caption{Results on TREC DL19 and DL20 datasets. In-domain supervised models DPR, ANCE and Contriever$^\text{FT}$ are trained on the MS-MARCO dataset and listed for reference. \textbf{Bold} indicates the best result across all models.}
\label{tab:dl}
\end{table*}

\begin{table*}[h]
\setlength{\abovecaptionskip}{0.2cm}
\setlength{\belowcaptionskip}{-0.0cm}
\centering
\small
\begin{tabular}{l|ccccccc}
\toprule
& Scifact & Arguana & Trec-Covid & FiQA    & DBPedia & TREC-NEWS & Avg. \\
\midrule
\multicolumn{7}{c}{nDCG@10} \\
\midrule
\multicolumn{7}{l}{\textit{w/o training}} \\
BM25       & 67.9    & 39.7 & 59.5 &  23.6 &  31.8   & 39.5 &43.7  \\
BM25+RM3 & 64.6    & 38.0  &  59.3   & 19.2  & 30.8    & 42.6 & 42.4\\
Contriever+HyDE     & 69.1    & \textbf{46.6}   & 59.3  &  27.3 &  36.8  &  44.0 & 47.2 \\
BM25+Q2D/PRF & \textbf{71.7} & 41.4 & 73.8 & 29.0 & 37.1 & 47.6 & \textbf{50.1}\\
BM25+KEQE & 70.5 & 40.7 & 66.6 & 22.0 & 38.8	& 48.3 & 47.8 \\
BM25+CSQE &  69.6   &  40.3  &  \textbf{74.2}    & 25.0 & 40.3  & \textbf{48.7} & 49.7 \\
\midrule
\multicolumn{7}{l}{\textit{reference. w/ training}} \\
DPR   & 31.8    & 17.5    & 33.2      & 29.5    & 26.3  & 16.1 & 25.7 \\
ANCE & 50.7    & 41.5    & 65.4 & 30.0    & 28.1    & 38.2 & 42.3\\
Contriever$^\text{FT}$ & 67.7 & 44.6 & 59.6 & \textbf{32.9} & \textbf{41.3} & 42.8 & 48.2 \\
\bottomrule
\end{tabular}
\caption{Results on low-resource retrieval datasets. \textbf{Bold} indicates the best result across all models.}
\label{tab:beir}
\end{table*}

%% file: experiments.tex
\section{Experiments}

\subsection{Setup}
\noindent \textbf{Datasets.}
Following~\citet{hyde}, we evaluate on (1) two web search datasets: TREC DL19~\cite{dl19} and TREC DL20~\cite{dl20}, which are based on the high-resource MS-MARCO dataset~\cite{msmarco}; and (2) six low-resource retrieval datasets from BEIR~\cite{beir} covering a variety of domains (e.g., medicine and finance) and query types (e.g., news headlines and arguments). 

\noindent \textbf{Baselines.}
We consider baselines from two categories: PRF methods and query expansion methods using LLMs. 
The PRF method we include is \textbf{BM25$+$RM3}~\cite{Lavrenko2001-lg,RM3}.
The query expansion methods with LLMs include:
(1) \textbf{Contriever+HyDE}, a KEQE method that employs hypothetical documents generated by LLMs to enhance unsupervised Contriever~\cite{contriever} model; (2) \textbf{BM25+GPR}~\cite{generative_relevance_feedback}, a query expansion method that applies PRF upon LLM-knowledge empowered hypothetical texts. GPR is a strong baseline that outperforms multiple SOTA PRF methods; (3) \textbf{BM25+Q2D/PRF}~\cite{prompt_qe}, a method that given initially-retrieved documents generates hypothetical documents instead of extracting key sentences from them; and (4) \textbf{BM25+KEQE}.

Moreover, we also include three supervised dense retrievers that are trained with over 500k human-labeled data of MS-MARCO for reference: (1) \textbf{DPR}; (2) \textbf{ANCE}, which involves sophisticated negative mining; and (3) \textbf{Contriever$^\text{FT}$}, which is the fine-tuned version of Contriever.

\noindent \textbf{Implementation.}
We utilize GPT-3.5-Turbo\footnote{We use the GPT-3.5-Turbo-0301 version. In our preliminary study, updating HyDE's LLM from Text-Davinci-003 to GPT-3.5-Turbo cannot improve results.} as our serving LLM for the trade-off between performance and cost. We sample from the LLM with a temperature of 1.0. BM25 retrieval and RM3 query expansion are performed using Pyserini~\cite{pyserini} with default hyper-parameters.
\ours utilizes the top-10 retrieved documents, with each truncated to at most 128 tokens, excluding the Arguana dataset where we keep the top-3 documents due to its lengthy passages. To increase diversity, for each API call, we sample N generations. For KEQE, $N=5$. As CSQE involves both KEQE
and corpus-originated expansions, we sample N = 2 for both KEQE and corpus-originated expansions, making only 4 generations total for fair comparison. The expanded query of each generation is further concatenated together to form the final query.

\subsection{Web Search Results}
Table~\ref{tab:dl} shows the retrieval results on TREC DL19 and DL20. \ours is able to bring a substantially larger improvement over BM25 compared to the strong PRF baseline RM3. Despite utilizing fewer LLM generations for expansion, CSQE surpasses KEQE on 5/6 metrics, showing the effectiveness of our corpus-steered approach. Moreover, CSQE consistently outperforms GPR on 5/6 metrics, which employs PRF on KEQE expansions, emphasizing the necessity of corpus-steered expansions. Comparing to Q2D/PRF, \ours shows superiority across all metrics. We interestingly find a phenomenon that if LLMs find no relevant documents in the initially-retrieved set, they will yield no expansions. However, in the case of Q2D/PRF, LLMs still need to generate documents, potentially being adversely affected by the presence of noisy documents~\cite{irrelevant_robust}. Without any training, CSQE with simple BM25 is able to beat the SOTA Contriever$^\text{FT}$ model across all metrics by a substantial margin.

\begin{table}[h]
\small
\centering
\footnotesize
\setlength{\tabcolsep}{3.5pt}
\setlength{\abovecaptionskip}{0.2cm}
\setlength{\belowcaptionskip}{-0.5cm}
\begin{tabular}{@{} l ccc @{}}
\toprule
Model & nDCG@1 & nDCG@5 & nDCG@10\\
\midrule
BM25 & 61.9 & 60.9 & 68.4 \\
BM25+KEQE & 50.0 & 48.7 & 62.0 \\
BM25+CSQE & 85.7 & 79.6 & 82.6 \\
RankGPT & 76.2 & 74.2 & 75.7 \\
\bottomrule
\end{tabular}
\caption{Results of CSQE on NovelEval. RankGPT refers to the GPT-3.5-Turbo-based reranker in \citet{chatgpt_rank}.}
\label{tab:csqe_on_contriever}
\end{table}

\subsection{Low-Resource Retrieval Results}
The results on 6 low-resource BEIR datasets are shown in Table~\ref{tab:beir}. Applying RM3 leads to performance drops on 5/6 datasets, while \ours is robust to domain shifts and is able to consistently improve BM25 on all datasets. Although KEQE can achieve similar results as Contriever$^\text{FT}$, CSQE is able to outperform both KEQE and Contriever$^\text{FT}$ by a large margin, demonstrating the strong generalizability of CSQE. \ours remains competitive when compared to Q2D/PRF, verifying the importance of corpus knowledge in low-resource scenarios.

%% file: analysis.tex
\section{Analysis}
\subsection{CSQE on Queries that LLMs Lack Knowledge}
To further verify that the reduction of hallucination leads to the performance improvements, we evaluate CSQE on NovelEval~\cite{chatgpt_rank}. NovelEval is a test set with queries and passages published after the
release of GPT-4, serving as a testbed where current LLMs have no knowledge and thus can only hallucinate. Following \citet{chatgpt_rank}, we report nDCG@1, nDCG@5, and nDCG@10. Interestingly, we find KEQE is not able to bring improvements while CSQE leads to remarkable improvements. Notably, BM25+CSQE outperforms a GPT-3.5-Turbo-based reranker which is more time-consuming to run, providing additional confirmation of the effectiveness of CSQE.

\subsection{CSQE on Dense Retrieval}
To test the versatility of \ours, we apply \ours on the unsupervised Contriever in Table~\ref{tab:csqe_on_contriever}. Following~\citet{hyde}, we encode each query expansion separately into dense embeddings and average their embeddings with the original query embedding as the final embedding. As the only difference between HyDE and KEQE on Contriever is their utilized LLMs (Text-Davinci-003 versus GPT-3.5-Turbo), we find they achieve similar results. Similar to the impact of \ours on BM25, \ours is able to improve Contriever significantly.
Interestingly, it is worth noting that in all cases, Contriever performs worse than BM25. 
Surprisingly, query expansion (Contriever+CSQE) is proven to be more effective than fine-tuning the model using 500K human-labeled data (Contriever$^\text{FT}$).

\begin{table}[h]
\small
\centering
\footnotesize
\setlength{\tabcolsep}{3.5pt}
\setlength{\abovecaptionskip}{0.2cm}
\setlength{\belowcaptionskip}{-0.0cm}
\begin{tabular}{@{} l ccc @{}}
\toprule
Model & mAP & nDCG@10 & Recall@1k\\
\midrule
Contriever &  24.0 & 44.5 & 74.6 \\
\quad +HyDE & 41.8 & 61.3 & 88.0 \\
\quad +KEQE & 41.7 & 62.2 & 87.4 \\
\quad +CSQE & 44.0 & 65.6 & 88.6  \\
\midrule
BM25 &  30.1 & 50.6 & 75.0 \\
\quad +KEQE &45.0 & 65.9 & 88.8  \\
\quad +CSQE & 47.6 & 68.6 & 89.0 \\
\midrule
Contriever$^\text{FT}$ & 41.7 & 62.1 & 83.6 \\
\bottomrule
\end{tabular}
\caption{Results of CSQE on Contriever on DL19.}
\label{tab:csqe_on_contriever}
\end{table}

\subsection{CSQE with Different LLMs}
We apply different LLMs for CSQE in Table~\ref{tab:diff_llms}. Utilizing Llama2-Chat-70B, we observe that BM25+CSQE outperforms MS-MARCO-tuned DPR, ANCE, and even Contriever$^\text{FT}$. However, a noticeable performance gap persists between Llama models and GPT-3.5-Turbo. Furthermore, we observe a consistent performance improvement with the increase in model size for both CSQE and KEQE. Across the models, CSQE consistently outperforms KEQE, verifying the effectiveness of CSQE. This conclusion also applies to DL20 with the exception that BM25+CSQE with Llama2-Chat-70B can not outperform but obtains comparable performance to the fine-tuned dense retrieval models, as outlined in Appendix~\ref{append:dl20}.

\begin{table}[h]
\small
\centering
\footnotesize
\setlength{\tabcolsep}{3.5pt}
\setlength{\abovecaptionskip}{0.2cm}
\setlength{\belowcaptionskip}{-0.0cm}
\begin{tabular}{l|ccc}
\toprule
Model & mAP & nDCG@10 & Recall@1k \\
\midrule
\multicolumn{4}{l}{\textit{BM25+CSQE}} \\
Llama2-Chat-7B & 39.1 &	59.9 & 86.5 \\
Llama2-Chat-13B & 41.4 & 61.6 & 86.1 \\
Llama2-Chat-70B & 43.6 & 63.4 & 87.6 \\
GPT-3.5-Turbo & 47.2 & 67.3 & 88.5 \\
\midrule
\multicolumn{4}{l}{\textit{BM25+KEQE}} \\
Llama2-Chat-7B & 37.5 & 57.1 & 82.0 \\
Llama2-Chat-13B & 38.8 & 58.3 & 83.7 \\
Llama2-Chat-70B & 40.9 & 59.3 & 84.9\\
GPT-3.5-Turbo & 45.0 & 65.9 & 88.8 \\
\midrule
\multicolumn{4}{l}{\textit{reference.}} \\ 
DPR & 36.5 & 62.2 & 76.9 \\
ANCE & 	37.1 & 64.5 & 75.5 \\
Contriever$^\text{FT}$ & 41.7 & 62.1 & 83.6 \\
\bottomrule
\end{tabular}
\caption{Results of CSQE with different LLMs on DL19.}
\label{tab:diff_llms}
\end{table}

\subsection{Case Studies}
Two qualitative examples of CSQE are shown in Table~\ref{tab:case_study}. As illustrated, KEQE has the tendency to hallucinate and introduce misleading unrelated keywords, while corpus-originated expansion in contrast is able to introduce meaningful and relevant texts to balance out KEQE's limitations.

\begin{table}[ht]
\setlength{\belowcaptionskip}{-0.5cm}
        \resizebox{0.95\columnwidth}{!}{
        \begin{tabular}{p{11cm}}
        \toprule
        
        \textbf{Query}: What is the screen resolution of vision pro?  \\
         \textbf{KEQE}: Vision pro is a \red{software product} from \red{zeiss} that is used for \red{data management} and evaluation in the field of \red{optometry}. However, since vision pro is a \red{software program, it doesn't have a screen resolution of its own.} Instead, the screen resolution you experience when using vision pro will depend entirely on the specifications of your computer hardware and display monitor\dots\\
         \textbf{Corpus-Originated Expansion}: The \green{extremely high-resolution displays} are one of the many features that set vision pro apart from its competitors. Vision pro uses \green{micro-oled} technology to pack \green{23 million pixels into two displays}, each the size of a postage stamp the vision pro display resolution is one of the many benefits of the device over its rivals the company refers to it as \green{"micro-oled"}, which has led some to mistake it for microled\dots\\
         
        \midrule
            \textbf{Question}: What are the new features of pytorch 2?  \\
         \textbf{KEQE}: Pytorch 2 brings a host of new features and improvements. Notably, it introduces a new feature called "\red{hybrid frontend}", which enables users to write complex models using both pytorch and \red{tensorflow} syntax. This allows for seamless integration of pytorch with other libraries and frameworks, offering users greater flexibility and control over their models\dots \\
         \textbf{Corpus-Originated Expansion}: Pytorch 2.0 has been released with fundamental changes to the way it \green{works at the compiler level, faster performance, and support for dynamic shapes and distributed}. The new release includes a \green{stable version of accelerated transformers}; and \green{torch.compile}, a feature that improves pytorch performance\dots \\
    
        \bottomrule
        \end{tabular}
        }
    \caption{Examples of CSQE on NovelEval. KEQE tends to produce non-factual and irrelevant texts, whereas Corpus-Originated Expansion introduces various meaningful and relevant texts. Certain expansions are omitted for the sake of space.}
    \label{tab:case_study}
    \end{table}

%% file: conclusion.tex
\section{Conclusion}
In this paper, we propose CSQE, which utilizes the relevance assessing ability of LLMs to balance out limitations associated with the intrinsic knowledge of LLMs. Experimental evaluation demonstrates CSQE's superiority over the LLM-knowledge empowered expansion methods and SOTA supervised Contriever$^\text{FT}$ model across various datasets.

%% file: limitations.tex
\section*{Limitations}
We acknowledge two limitations in our work: computational overhead and reliance on closed-source models. The utilization of OpenAI LLMs necessitates API calls, resulting in increased processing time and latency. However, in retrieval tasks where latency is less crucial, such as legal case retrieval, our method may offer benefits. Moreover, our approach does not necessitate training, making it more accessible to researchers and practitioners without extensive GPU resources. Additionally, the unavailability of the LLMs' source models and training data restricts our ability to conduct thorough analysis. There may exist data contamination issues~\cite{data_contamination} where some of our test examples are already present in the training data of the LLMs.

We utilized ChatGPT to correct the grammar in our paper and ensured that none of the text was directly generated by ChatGPT.

%% file: acknowledgement.tex
\section*{Acknowledgement}
This research was supported by the Hybrid Intelligence Center, a 10-year program funded by the Dutch Ministry of Education, Culture and Science through the Netherlands Organisation for Scientific Research, \url{https://hybrid-intelligence-centre.nl}, and project VI.Vidi.223.166 of the NWO Talent Programme which is (partly) financed by the Dutch Research Council (NWO).

%% file: appendix.tex
\onecolumn
\appendix

\section{Appendix}
\label{sec:appendix}
\subsection{Instruction of \ours}
\label{instruc:prf-llm}
\begin{tcolorbox}
Query: "how are some sharks warm blooded"

Retrieved documents:

1. Most sharks are cold-blooded. Some, like the Mako and the Great white shark, are partially warmblooded (they are endotherms). Cold blooded although if you've ever seen a Great White Shark hunt sea lions you'd be thinking they would have to be hotblooded. Actually the Salmon Shark is a warm blooded shark.

2. Are sharks cold-blooded or warm-blooded? Sharks have a reputation as cold-blooded and despite how negative that term is, it is not entirely inaccurate. Sharks are by no means evil, vicious killers like that quote suggests. Nonetheless, sharks are, for the most part anyways, efficient ectothermic predators. Endo vs Ecto.

3. Great white sharks are some of the only warm blooded sharks. This allows them to swim in colder waters in addition to warm, tropical waters. Great White sharks can be found asâ¦ north as Alaska and as south as the southern tip of South America. They exist worldwide, everywhere in-between. 5 people found this useful.

4. Sharks' blood gives them turbo speed. Several species of shark and tuna have something special going on inside their bodies. For a long time, scientists have known that some fish species appear warm-blooded. Salmon sharks can elevate their body temperatures by up to 20 degrees compared to the surrounding water, for example.

You will begin by examining the initially retrieved documents and identifying the ones that are relevant, even partially, to the query. Once the relevant documents are identified, you will extract the key sentences from each document that contribute to their relevance.
\\ \\
Based on the query "how are some sharks warm blooded", I have examined the initially retrieved documents. Here are the relevant documents and the key sentences extracted from each: 

Document 1:\\
"Most sharks are cold-blooded. Some, like the Mako and the Great white shark, are partially warm-blooded (they are endotherms)."

"Actually, the Salmon Shark is a warm-blooded shark."

Document 3:\\
"Great white sharks are some of the only warm-blooded sharks."

"This allows them to swim in colder waters in addition to warm, tropical waters."

Document 4:\\
"Salmon sharks can elevate their body temperatures by up to 20 degrees compared to the surrounding water, for example."\\

Query: "$\{q\}$"

Retrieved documents:

1. $\{d_1\}$

2. $\{d_2\}$

\ldots

$\{k\}$. $\{d_k\}$

You will begin by examining the initially retrieved documents and identifying the ones that are relevant, even partially, to the query. Once the relevant documents are identified, you will extract the key sentences from each document that contribute to their relevance.
\end{tcolorbox}

\newpage
\subsection{Dataset Statistics}
Details about the retrieval datasets are shown in Table~\ref{tab:data_stat_retrievaaaal}.

\begin{table}[hbt]
\setlength{\abovecaptionskip}{0.1cm}
\setlength{\belowcaptionskip}{-0.2cm}
\centering
\small
\setlength{\tabcolsep}{3pt}
\begin{tabular}{@{} l rr @{}}
\toprule
Dataset & \#Test & \#Corpus \\
\midrule
DL19 & 43 & 8,841,823 \\
DL20 & 50 & 8,841,823 \\
Scifact & 300 & 5183 \\
Arguana & 1406 & 8674 \\
Trec-Covid  & 50 & 171,332 \\
FiQA   & 648 & 57,638 \\
DBPedia    & 400 & 4,635,922 \\
TREC-NEWS & 57 & 594,977 \\
NovelEval & 21 & 420 \\
\bottomrule
\end{tabular}
\caption{Dataset Statistics}
\label{tab:data_stat_retrievaaaal}
\end{table}

\subsection{CSQE with Different LLMs on DL20}
\label{append:dl20}
\begin{table}[h]
\small
\centering
\footnotesize
\setlength{\tabcolsep}{3.5pt}
\setlength{\abovecaptionskip}{0.2cm}
\setlength{\belowcaptionskip}{-0.0cm}
\begin{tabular}{l|ccc}
\toprule
Model & mAP & nDCG@10 & Recall@1k \\
\midrule
\multicolumn{4}{l}{\textit{BM25+CSQE}} \\
Llama2-Chat-70B & 41.4 & 61.5 & 86.5 \\
GPT-3.5-Turbo & 46.5 & 66.2 & 89.1 \\
\midrule
\multicolumn{4}{l}{\textit{BM25+KEQE}} \\
Llama2-Chat-70B & 42.0 & 58.5 & 85.2\\
GPT-3.5-Turbo &	42.8 & 60.5 & 88.3\\
\midrule
\multicolumn{4}{l}{\textit{reference.}} \\ 
DPR & 41.8 & 65.3 & 81.4 \\
ANCE & 	40.8 & 64.6 & 77.6 \\
Contriever$^\text{FT}$ & 43.6 & 63.2 & 85.8 \\
\bottomrule
\end{tabular}
\caption{Results of CSQE with different LLMs on DL20.}
\label{tab:diff_llms_dl20}
\end{table}